\pdfoutput=1
\documentclass{article}
\usepackage{amsmath}
\usepackage{amsthm}
\usepackage{authblk}
\usepackage{graphicx}
\usepackage[numbers]{natbib}
\usepackage{hyperref}
\usepackage[margin=1.2in]{geometry}

\title{Outlier Impact: Detection by Consequences}
\author{Daniel Ting}
\author{Ilya Gorbachev}
\author{Sammy Shen}
\affil{Meta \\ {\tt\{dting, ilyagorbachev, sammyshen\}@meta.com}}

\date{}

\begin{document}

\maketitle

\section{Introduction}

Experimentation heavily relies on the Central Limit Theorem (CLT) to obtain a reliably correct difference-in-means test. 
However, the asymptotic nature of the CLT means that it cannot always be applied in practice. One major reason is due to the presence of outliers. 
While there are many methods for flagging outliers, they are typically selected based on how "unusual" a point is.
Indeed the definition quoted by surveys \cite{boukerche2020outlier}
and attributed to Grubbs~\cite{grubbs1969procedures} is: 
“an outlier is one that appears to deviate markedly from other members of the sample in which it occurs.”

This leads statistical approaches surveyed in~\cite{chandola2009anomaly}, for example,  to flag outliers based on the estimated probability for a point to appear. These approaches are used in industry. Both Optimizely~\cite{optimizely_outliers} and Dynamic Yield~\cite{dynamicyield_outliers} use a normality based Grubbs test to flag outliers that are 3 standard deviations away from the mean. However, this criterion is tangential to the objectives of an experimenter, whose main concern is 1) whether the estimate of the effect size is reliable, and 2) if the uncertainty in the estimate is properly quantified. These are needed to make decisions and assess their risk. 

We propose a method that directly addresses experimenters' concerns by flagging and excluding outliers based on their impact on a hypothesis test and its validity. Furthermore, our method is nearly assumption-free. It relies solely on the experiment randomization, a weak assumption that the CLT is a reasonable approximation after outliers are removed, and a mild assumption that treatment effects are small. In other words, we rely only on assumptions that are needed for experimentation to be useful in the first place. The main idea is to examine how much individual data points can influence the estimated Average Treatment Effect (ATE), and the False Positive Rate (FPR) or Type I error of a test. 
If the resulting FPR does not match the nominal FPR set by the experimenter, then we flag points as outliers.

\section{Methodology}
We are interested in a standard difference-in-means z-test and flagging and excluding outliers that jeopardize the validity of the test. 

Our methodology consists of 5 parts:
\begin{enumerate}
\item Randomly bin points so they are close to Normally distributed.
\item Sort to identify candidate outlier sets.
\item Construct a ground truth FPR with A/A experiments.
\item Construct a robust test excluding outlier candidates.
\item Estimate the FPR when candidate outliers are included.
\end{enumerate}

For the first part, the original $n$ points are randomly grouped into $g \gg 1$ bins, typically via a hash function, and summed or averaged. The purpose of this step is to make each input point for our flagging method approximately normal if it doesn't contain an outlier. 
The purpose of this step is to ensure the "points" that we consider are approximately normal. While this step is not strictly necessary, it allows us to avoid making any parametric assumptions. Having some distribution is necessary as we will use the distribution to impute points to replace candidate outliers. 
This also means that instead of flagging individual points, we flag a small group of points that contain an outlier. Note that excluding a small group of random points (e.g. $1/1000^{th}$ of the data) from an experiment has minimal impact on its power. Henceforth, we refer to these binned points as simply points.

For the second part, we simply sort the points. Rather than considering individual points as outliers, we determine whether the top-k points should be considered outliers. 

Typically, a large part of the difficulty in flagging outliers is the lack of a ground truth. 
By constructing hypothetical A/A experiments where the experimenter sets the nominal FPR or Type I error, we establish a ground truth. 
We do this by simply pooling all the data in treatment and control groups. This allows us to construct a test of the same size without needing to reuse data points, and is more faithful to the true randomization procedure. When the treatment effect of an experiment is small, then LeCam's third~lemma \cite{vandervaart1998asymptotic} gives that the variance (and hence the A/A experiment itself) has the same asymptotic behavior as an A/A experiment where everyone received the control treatment.

The fourth, and most important, part of our methodology is to take a standard z-test with outlier points included and compare it to a robust version of the test that excludes outliers. In particular, we replace the candidate outlier points $x^{(n-i)}$ with imputed points $\tilde{x}^{(n-i)}$ that are more robust estimates of what the top $k$ values should be. From the perspective of a z-test, outliers behave like true positive effects. Hence, we can estimate the false added effect when there are $k$ outliers as
\begin{align}
    \Delta_{False}^{(k)} &= \frac{1}{n}\sum_{i=1}^k (x^{(n-i)} - \tilde{x}^{(n-i)})
\end{align}
The estimated FPR from including outliers is thus the probability that a test with desired FPR $\alpha$ detects a statistically significant result when the  effect size is this $\Delta_{False}^{(k)}$. In other words, it is the power of a the test when plugging in $\Delta_{False}^{(k)}$ as the true effect.

To construct this test, we first obtain a robust estimator of the variance that excludes outliers. 
This gives an estimate $\hat{\sigma}_{\mathrm{k-robust}}(\mathcal{D})$ that quantifies the variability of the mean for data $\mathcal{D}$ and yields a z-test statistic: 
\begin{align}
T(\mathcal{D}) &= \hat{\mu}(\mathcal{D}) / \hat{\sigma}_{\mathrm{k-robust}}(\mathcal{D}).
\end{align}
Secondly, since we wish to compare the impact of the candidate outliers, we construct another dataset $\tilde{\mathcal{D}_k}$ which imputes and replaces the top-k order statistics with their predicted values $\tilde{x}^{(n-i)} = \hat{F}^{-1}(1-i/(n+1))$ where $\hat{F}$ is the estimated normal CDF that uses our robust estimate of the variance. 

Our final comparison computes how much the $k$ candidate outliers can increase the FPR and assigns an outlier score to these points.
\begin{align}
    \widehat{FPR}_k^{increase} &= P\left(Z + \frac{\Delta^{(k)}_{False}}{\hat{\sigma}_{k-robust}(\mathcal{D})}  > z_\alpha\right) - \alpha.
\end{align}
Here, $Z \sim Normal(0,1)$, $\alpha$ is the desired significance level of a one-sided test, and $z_\alpha = \Phi^{-1}(1-\alpha)$ is the threshold used for a one-sided z-test with significance $\alpha$.

\section{Interpretation, properties, and result}
\begin{figure}
    \includegraphics[width=\textwidth, height=1.7in]{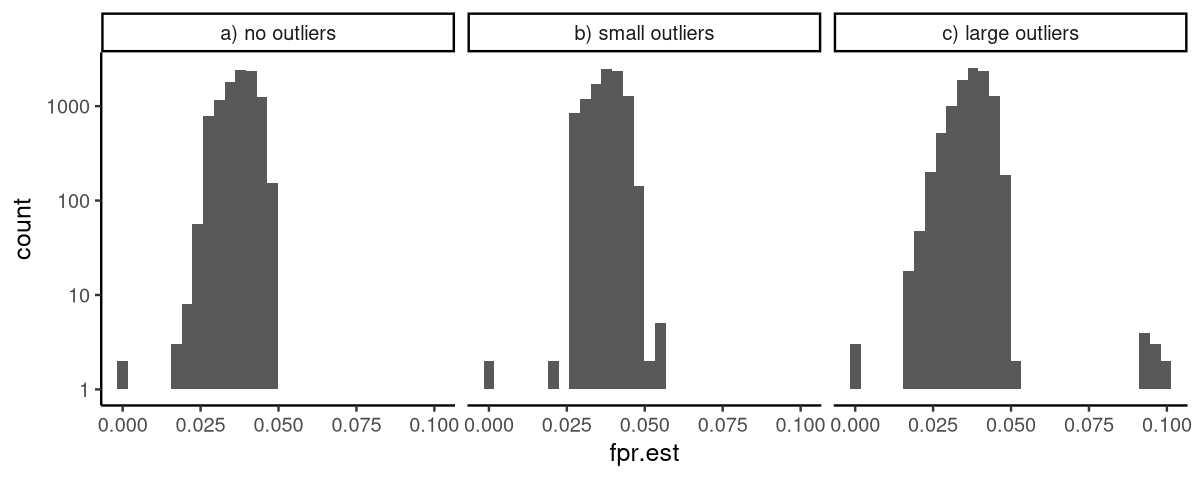}
    \caption{Distribution of outliers scored by FPR estimate.} 
    \label{fig:simulation}
\end{figure}

Setting a threshold based on impact to the FPR enables us to flag only results that make a practical difference to experimenters. 
This avoids a common problem where a point may be "unusual" and flagged by some other outlier detection method, but it's impact to the final estimate of interest is small. For example, for an experiment with 1 million approximately normally distributed points, a point that is 10 standard deviations away from the mean is extremely unusual. The probability of observing a more extreme value is $< 10^{-23}$. However, its impact to the mean is a tiny $10^{-5}$ and it increases the z-test statistic by just $10^{-5} / 10^{-3} =0.01$. In other words, the "outlier" makes almost no difference to the hypothesis test. Figure \ref{fig:simulation} illustrates our method in simulation with a smaller $n=1000$ and outliers of different sizes.
Scores close to or smaller than the nominal FPR of $0.05$ indicate a point is not an outlier that impacts the test.
When there are no true outliers, there are no collections of point that would pull the FPR above the nomial 0.05. In other words, the estimated false effect $\Delta_{False}$ is close to 0 or smaller. 
With modest, but extremely rare outliers 6 standard deviations away, the outliers only drag the FPR up slightly. With large outliers (20 standard deviations away), the FPR drastically increases.

In hypothesis testing based experimentation, experimenters make decisions based on the statistical significance, and they express preferences about how to make decisions through the one parameter they control in the test: the significance level or the FPR. 
Our method allows experimenters to express their tolerance of outliers through this parameter. For example, even if the nominal FPR is set to 5\%, experimenters may decide that a test with a 7\% false positive rate is acceptable. In the outlier example above, a 0.01 increase in the z-statistic is equivalent to increasing the FPR to 5.1\%. 

\bibliographystyle{plain}
\bibliography{references}

@article{chandola2009anomaly,
  title={Anomaly detection: A survey},
  author={Chandola, Varun and Banerjee, Arindam and Kumar, Vipin},
  journal={ACM computing surveys (CSUR)},
  volume={41},
  number={3},
  pages={1--58},
  year={2009},
  publisher={ACM New York, NY, USA}
}

@article{boukerche2020outlier,
  title={Outlier detection: Methods, models, and classification},
  author={Boukerche, Azzedine and Zheng, Lining and Alfandi, Omar},
  journal={ACM Computing Surveys (CSUR)},
  volume={53},
  number={3},
  pages={1--37},
  year={2020},
  publisher={ACM New York, NY, USA}
}

@article{grubbs1969procedures,
  title={Procedures for detecting outlying observations in samples},
  author={Grubbs, Frank E},
  journal={Technometrics},
  volume={11},
  number={1},
  pages={1--21},
  year={1969},
  publisher={Taylor \& Francis}
}

@misc{dynamicyield_outliers,
  title = {Outliers detection – ridding the extreme events threatening your A/B tests},
  author = {Gidi Vigo},
  howpublished = {\url{https://www.dynamicyield.com/lesson/outliers-detection/}},
  note = {Accessed: 2024-09-01}
}

@misc{optimizely_outliers,
  title = {How Optimizely Experimentation handles outliers},
  howpublished = {\url{https://support.optimizely.com/hc/en-us/articles/4410289414413-How-Optimizely-Experimentation-handles-outliers}},
  note = {Accessed: 2024-09-01}
}

@book{vandervaart1998asymptotic,
  title = {Asymptotic {{Statistics}}},
  author = {{van der Vaart}, Aad W.},
  year = {1998},
  month = jan,
  volume = {3},
  publisher = {{Cambridge University Press}},
  address = {{Cambridge, UK}},
  isbn = {0-521-49603-9 0-521-78450-6},
}

\end{document}